\documentclass[aps,prb,preprint,showpacs]{revtex4}
\usepackage{natbib}
\usepackage{epsfig} 
\begin{document}
\title{Four electrons in a two-leg Hubbard ladder: Exact ground states.}
\author{Endre~Kov\'acs and Zsolt~Gul\'acsi}  
\affiliation{Department of Theoretical Physics, University of Debrecen, 
H-4010 Debrecen, Hungary}
\date{\today}
\begin{abstract}
In the case of a two-leg Hubbard ladder
we present a procedure which allows the exact deduction of the ground state
for the four particle problem in arbitrary large lattice system, in a 
tractable manner, which involves only a reduced Hilbert space region 
containing the ground state. In the presented case, the method leads
to nine analytic, linear, and coupled equations providing the ground 
state. The procedure which is applicable to few particle problems and other 
systems as well is based on an ${\bf r}$-space representation of the wave 
functions and construction of symmetry adapted orthogonal basis wave vectors 
describing the Hilbert space region containing the ground state. Once the
ground state is deduced, a complete quantum mechanical characterization of the
studied state can be given. Since the analytic structure of the ground state
becomes visible during the use of the method, its importance is not reduced
only to the understanding of theoretical aspects connected to exact 
descriptions or potential numerical approximation scheme developments, but is
relevant as well for a large number of potential technological application
possibilities placed between nano-devices and quantum calculations, where
the few particle behavior and deep understanding are important key aspects
to know. 
\end{abstract}
\maketitle

\section{Introduction}
In condensed matter context, in experiments or related theoretical 
interpretations we often encounter small number of particles confined in a 
system or device, for example, in the case of
quantum dots \cite{i1}, quantum well structures \cite{i2}, mesoscopic 
systems \cite{d5},  experimental entanglement 
\cite{i5}, micro-crystals \cite{i5x},
could gases trapped in optical lattices \cite{i6,i60}, 
optical bound states \cite{i6a}, segregation \cite{i6b},
interfacial stress and fracture \cite{i6c}, self organized structures 
\cite{i6d}, sintering \cite{i6e}, or compounds 
studied in the low concentration limit \cite{i7}. 
Such problems, presenting both theoretical \cite{i7a,i7b,i7c,f1}
and technological \cite{i6b,i6c,i6e,i6aa,i6bb,i6cc} interest have
continuously attracted increasing attention. Starting from even one 
electron problems solved exactly \cite{f1}, several cases of interest for two
\cite{f2,f2a,f2b,f2c,f2d,f2e}, three \cite{i7a,i7b,i6dd,i6ee,i6eex}, 
four \cite{d2,d2a,d2b}, 
or few \cite{f5a,f5b,f5c} particles have 
been studied, concentrating on the model behavior in the low concentration 
limit, or motivated by experimentally measured characteristics. 
In this hierarchy of the 
increasing number of carriers in the study of a given problem, the particle 
number four ($N_p=4$) represents a special case, since it is close to the 
particle 
number limit around which one can hope that deep rigorous descriptions can be 
made \cite{m01} even in the non-integrable cases, the problem is also 
treatable from the numerical side as well \cite{m02}, statistics and $T \ne 0$
characterization can be given \cite{m03}, and the problem retain even 
many-body aspects of the system's behavior \cite{m1,m2,m3}. 

The simulations on the $N_p=4$ particle problem started more than a decade
ago \cite{m02,m4}, but up today, only few valuable results are known in this 
subject in the condensed matter context, as follows. The energy dependence of 
the maximal 
Lyapunov exponent has been studied for 1D Lenard-Jones system \cite{d1}, the 
spinless fermion case has been analyzed as a simplified model for correlated 
electrons \cite{d2,d2a}, the conjecture of the Andreev-Lifshitz supersolid 
has been studied \cite{d2b}, entangled states have been described in the high 
frequency region \cite{d3}, doped quantum well structures have been
investigated  \cite{i2}, special cases where only two pairs of particles 
interact on a lattice were considered \cite{d4}, localization lengths have
been estimated in 1D disordered systems \cite{d5}, and the behavior in the 
presence of Coulomb forces has been analyzed \cite{d4a}.
As can be seen, the knowledge accumulated in this direction is relatively poor.
Approximated procedures have been applied in different conditions for
different systems of interest, but the level of exact characteristics
has not been reached yet.

The need to study at exact level system holding $N_p=4$ particles is 
enhanced by several motivations. First of all, $N_p=4$ it is placed in the 
low density limit, and 
as known, in this limit, especially in low dimensions, no class of diagrams 
can be neglected in describing the system \cite{uu1}. Given by this 
difficulty, one often finds that traditional approximation schemes which work 
at higher densities, here fail \cite{uu2} or provide unphysical results 
\cite{uu3}. Secondly, we are placed in the concentration limit
where the formation of Fermi liquid properties can be studied \cite{d2}, and
since this parameter region is usually numerically accessible, research with 
analytical focus, starting from numerical results, also can be done. Thirdly, 
as several times has been accentuated \cite{uu4,uu5}, key aspects of the 
unapproximated 
descriptions are often hidden in the few particle cases. The four particle 
case seems to be tractable also from this point of view.
Finally, in $N_p=4$ case we face a situation which experimentally is produced,
having potential application possibilities in several areas,
as for example in the study of entangled 
states \cite{uu6}, non-local character of quantum theory \cite{uu6a}, 
high precision spectroscopy \cite{uu6b}, quantum communication, quantum 
cryptography, and quantum computation \cite{uu6c}, fields where deep and
high quality results are clearly demanded \cite{uu7}.

Starting from the motivations presented above, we show in this paper that
for the $N_p=4$ case, exact, analytical, and explicit results 
holding essential information about the system behaviour can be indeed 
provided, even for arbitrary large systems. To show this, we present below
the exact ground state for four interacting electrons placed in  
an arbitrary large two leg Hubbard ladder described by periodic boundary 
conditions. This is given in conditions for which,
even the known three (quantum mechanical) particle exact results are very rare 
for systems taken outside of one dimension (see Ref.[\cite{i6eex}] and the 
references therein), hence we hope that the presented results will generate
creative advancements.

In order to obtain such results, a direct space representation is used
for the wave functions. Starting from local particle configurations, symmetry 
adapted ortho-normalized basis wave vectors are constructed. Based on these,
in the studied case, an explicit and analytic closed system of 9 equations is 
constructed, whose secular equation provides the ground state wave function 
and energy. 
Deducing the ground state wave function for different microscopic parameters
of the model, ground state expectation values are calculated for different
physical quantities of interest, and correlation functions are deduced in 
order to characterize the ground state properties.  

The method which is described here is in principle not model or particle 
number dependent, and could be applied for other systems as well. In 
presenting our calculations, the
aim was not to hide the obtained results behind a numerical treatment or 
deduced symmetry properties, (which certainly also can be done), but to 
show clear, visible, and explicit properties which, based on the provided
essential characteristics, could enhance further creative thinking or
applications.
In order to underline the importance of these aspects we note for example,
that in recent studies made for states containing 2-4 particles,
especially in attempts to characterize the entanglement \cite{uu6}, 
or quantum dots \cite{uu6aa}, often
the analysis must be made without to know the state completely \cite{uu7,uu8}.
We show below how such ingredients, at least at the level of the ground state,
are possible to overcome. 

The remaining part of the paper is structured as follows. Section II. 
presents the Hamiltonian, the deduction procedure and the ground state wave 
functions, Section III. exemplifies the physical properties of the ground 
state, Sect. IV. presents the summary and conclusions of the paper, while the 
Appendices A - B presenting mathematical details, close the presentation.

\section{Hamiltonian and ground state wave functions.}

The strategy which we use for presentation is the following one.
We have chosen a simple model which allow us to characterize the construction 
of exact ground states in the presence of four particles. After presenting the
results we indicate how the procedure could be applied for other systems as 
well.

\subsection{Presentation of the Hamiltonian}

The Hamiltonian we use for presentation has the form of a standard two-leg 
Hubbard ladder Hamiltonian
\begin{eqnarray}
\hat H = - t_{\parallel} \sum_{<i,j>_{\parallel},\sigma} (\hat c^{\dagger}_{i,
\sigma} \hat c_{j,\sigma} + H.c.) - t_{\perp} \sum_{<i,j>_{\perp},\sigma}
(\hat c^{\dagger}_{i,\sigma} \hat c_{j,\sigma} + H.c.)
+U \sum_{i} \hat n_{i,\uparrow} \hat n_{i,\downarrow} ,
\label{ez1}
\end{eqnarray}
where $\hat c^{\dagger}_{i,\sigma}$ creates an electron at site $i$ with spin
$\sigma$, $t_{\alpha}$ holding the index $\alpha=\parallel, \perp$ are 
nearest-neighbor hopping amplitudes along and perpendicular to ladder legs,
$U$ is the on-site Coulomb interaction, and $<i,j>_{\alpha}$ represents 
nearest-neighbor sites in $\alpha$ direction taken into account in the sum 
over the lattice sites only once.

\subsection{The construction of the basis wave vectors}

If we would like to analyze by exact diagonalization the four particle problem
in the singlet case in a two leg Hubbard  ladder containing $N$ lattice sites,
we must treat numerically a Hilbert space of $d_H=[N(N-1)/2]^2$ dimensions, 
where for example at $N=30$ we have $d_H = 2.16 \cdot 10^5$, and for 
$N\to \infty$ one encounters $d_H \to \infty$ as $d_H \sim N^4$.

We show below how it is possible to deduce exactly the ground state for a 
such type of system in the case of an arbitrary large two leg Hubbard ladder 
based on only nine linear and analytic equations, and to extract essential 
information from the obtained results. In order to do this, first of
all we delimit exactly the Hilbert space region (${\cal{H}}_g$) containing the
ground state by the construction of nine type of orthogonal basis wave vectors
spanning ${\cal{H}}_g$. This procedure is presented below. 

\subsubsection{The generating configurations}

We are interested first to have an image about the possible type of states
of the studied four particles in the system under consideration.
To obtain such type of information, we number all lattice sites of the ladder 
as shown
in Fig.1 (periodic boundary conditions are considered). In the figure,
$N$, considered even number, denotes the number of sites within the system, 
while $n=N/2$ gives the number of rungs, respectively. Using now an 
${\bf r}$-space representation, one observes that since the ladder legs, and
the spin reversed configurations are equivalent, the studied four particles
can be placed into the system only in nine possible ways, as depicted in Fig.2.
The presented possibilities, denoted by capital letters $A$ to $J$ will provide
nine type of basis wave vectors (denoted by the same letters), whose 
construction is presented below.
We mention that the subscripts $i,j,k$ are denoting particle positions
within the considered states $A$ to $J$ presented in Fig.2, which are
such chosen, to have the first particle position placed into the origin (e.g. 
lattice site 1).  
In the following, the nine possible four-particle states
presented in Fig.2 will be called {\it generating configurations}. How one
arrives from the generating configuration $X=A,B,...,J$ to the base
vector $|X\rangle$, is explained in the following two subsections.

\subsubsection{The sum of configurations related to each generating 
configuration}

Each generating configuration provides other seven related configurations 
(brother configurations) of the same type. These are obtained by
a) rotating the generating configuration by 180 degrees along the
longitudinal symmetry axis of the ladder, b) rotating the generating 
configuration by 180 degrees along the symmetry axis perpendicular to the 
ladder, c) rotating by 180 degrees the configuration obtained at b) along the 
longitudinal symmetry axis of the ladder, and finally, d) other four related
configurations are obtained by reversing all spin orientations in the 
generating configuration and the configurations deduced at points a)-c).
As an example, the eight related configurations describing the 
state $D_{i,j}$ taken at $i=2, j=3$, are depicted in the first column of Fig.3.

After this step, since all lattice sites are equivalent, the different 
,,related'' configurations are translated by {\it elementary translation}
$N/2$ times along the ladder, and
all the contributions are added. We obtain in this manner a sum of 
configurations for each generating configuration. Such a sum contains 
$8 \times N/2$ components. For example, in the 
case of the $D_{2,3}$ state, this sum is presented in Fig.3.

The procedure described above must be effectuated separately for each 
generating configuration. As a result, we obtain at this point nine 
configuration sums. Each of these sums will give rise to one basis wave vector
as follows.

\subsubsection{The basis wave vectors}

A given configuration sum described in the previous subsection provides one 
basis wave vector if each individual configuration of the sum is written in 
mathematical form via four creation operators acting on the bare vacuum.
In order to do this, we have to fix the order of creation operators for each
type of contribution, which has been done as follows. For two doubly occupied
sites we write the creation operators of the couples next to each other, first
the spin up, then the spin down contribution, as $\hat c^{\dagger}_{i,\uparrow}
\hat c^{\dagger}_{i,\downarrow}\hat c^{\dagger}_{j,\uparrow}
\hat c^{\dagger}_{j,\downarrow}|0\rangle$, where only the restriction 
$i\ne j$ exists. In the case of basis wave vectors containing only one doubly 
occupied site at $i$ one uses $\hat c^{\dagger}_{i,\uparrow}
\hat c^{\dagger}_{i,\downarrow}\hat c^{\dagger}_{j,\uparrow}
\hat c^{\dagger}_{k,\downarrow}|0\rangle$, were $i \ne j$ and $i \ne k$ must 
hold. Finally, for cases without double occupancies, the convention
$\hat c^{\dagger}_{i,\uparrow} \hat c^{\dagger}_{j,\uparrow}
\hat c^{\dagger}_{k,\downarrow} \hat c^{\dagger}_{l,\downarrow}|0\rangle$
is considered, where $i < j$, and $k < l$ must hold. Using these conventions,
for example, in the case of $|D_{i,j}\rangle$, taken at $i=2,j=3$ and
depicted in Fig.3. the result becomes
\begin{eqnarray}
|D_{2,3}\rangle =
&&((\hat{c}_{1\uparrow}^{\dagger}\hat{c}_{1\downarrow}^{\dagger}
\hat{c}_{2\uparrow}^{\dagger}\hat{c}_{(n+3)\downarrow}^{\dagger}
+\hat{c}_{2\uparrow}^{\dagger}\hat{c}_{2\downarrow}^{\dagger}
\hat{c}_{3\uparrow}^{\dagger}\hat{c}_{(n+4)\downarrow}^{\dagger}
+\hat{c}_{3\uparrow}^{\dagger}\hat{c}_{3\downarrow}^{\dagger}
\hat{c}_{4\uparrow}^{\dagger}\hat{c}_{(n+5)\downarrow}^{\dagger}
+\dots)
\nonumber\\
&&+(\hat{c}_{1\uparrow}^{\dagger}\hat{c}_{1\downarrow}^{\dagger}
\hat{c}_{(n+3)\uparrow}^{\dagger}\hat{c}_{2\downarrow}^{\dagger}
+\hat{c}_{2\uparrow}^{\dagger}\hat{c}_{2\downarrow}^{\dagger}
\hat{c}_{(n+4)\uparrow}^{\dagger}\hat{c}_{3\downarrow}^{\dagger}
+\hat{c}_{3\uparrow}^{\dagger}\hat{c}_{3\downarrow}^{\dagger}
\hat{c}_{(n+5)\uparrow}^{\dagger} \hat{c}_{4\downarrow}^{\dagger}
+\dots)
\nonumber\\
&&+(\hat{c}_{(n+1)\uparrow}^{\dagger}\hat{c}_{(n+1)\downarrow}^{\dagger}
\hat{c}_{(n+2)\uparrow}^{\dagger}\hat{c}_{3\downarrow}^{\dagger}
+\hat{c}_{(n+2)\uparrow}^{\dagger}\hat{c}_{(n+2)\downarrow}^{\dagger}
\hat{c}_{(n+3)\uparrow}^{\dagger}\hat{c}_{4\downarrow}^{\dagger}
+\hat{c}_{(n+3)\uparrow}^{\dagger}\hat{c}_{(n+3)\downarrow}^{\dagger}
\hat{c}_{(n+4)\uparrow}^{\dagger}\hat{c}_{5\downarrow}^{\dagger}
+\dots)
\nonumber\\
&&+(\hat{c}_{(n+1)\uparrow}^{\dagger}\hat{c}_{(n+1)\downarrow}^{\dagger}
\hat{c}_{3\uparrow}^{\dagger} \hat{c}_{(n+2)\downarrow}^{\dagger}
+\hat{c}_{(n+2)\uparrow}^{\dagger}\hat{c}_{(n+2)\downarrow}^{\dagger}
\hat{c}_{4\uparrow}^{\dagger} \hat{c}_{(n+3)\downarrow}^{\dagger}
+\hat{c}_{(n+3)\uparrow}^{\dagger}\hat{c}_{(n+3)\downarrow}^{\dagger}
 \hat{c}_{5\uparrow}^{\dagger} \hat{c}_{(n+4)\downarrow}^{\dagger}
+\dots)
\nonumber\\
&&+(\hat{c}_{3\uparrow}^{\dagger}\hat{c}_{3\downarrow}^{\dagger}
\hat{c}_{2\uparrow}^{\dagger}\hat{c}_{(n+1)\downarrow}^{\dagger}
+\hat{c}_{4\uparrow}^{\dagger}\hat{c}_{4\downarrow}^{\dagger}
\hat{c}_{3\uparrow}^{\dagger}\hat{c}_{(n+2)\downarrow}^{\dagger}
+\hat{c}_{5\uparrow}^{\dagger}\hat{c}_{5\downarrow}^{\dagger}
\hat{c}_{4\uparrow}^{\dagger}\hat{c}_{(n+3)\downarrow}^{\dagger}
+\dots)
\nonumber\\
&&+(\hat{c}_{3\uparrow}^{\dagger}\hat{c}_{3\downarrow}^{\dagger}
\hat{c}_{(n+1)\uparrow}^{\dagger} \hat{c}_{2\downarrow}^{\dagger}
+\hat{c}_{4\uparrow}^{\dagger}\hat{c}_{4\downarrow}^{\dagger}
\hat{c}_{(n+2)\uparrow}^{\dagger} \hat{c}_{3\downarrow}^{\dagger}
+\hat{c}_{5\uparrow}^{\dagger}\hat{c}_{5\downarrow}^{\dagger}
\hat{c}_{(n+3)\uparrow}^{\dagger} \hat{c}_{4\downarrow}^{\dagger}
+\dots)
\nonumber\\
&&+(\hat{c}_{(n+3)\uparrow}^{\dagger}\hat{c}_{(n+3)\downarrow}^{\dagger}
\hat{c}_{(n+2)\uparrow}^{\dagger}\hat{c}_{1\downarrow}^{\dagger}
+\hat{c}_{(n+4)\uparrow}^{\dagger}\hat{c}_{(n+4)\downarrow}^{\dagger}
\hat{c}_{(n+3)\uparrow}^{\dagger}\hat{c}_{2\downarrow}^{\dagger}
+\hat{c}_{(n+5)\uparrow}^{\dagger}\hat{c}_{(n+5)\downarrow}^{\dagger}
\hat{c}_{(n+4)\uparrow}^{\dagger}\hat{c}_{3\downarrow}^{\dagger}
+\dots)
\nonumber\\
&&+(\hat{c}_{(n+3)\uparrow}^{\dagger}\hat{c}_{(n+3)\downarrow}^{\dagger}
\hat{c}_{1\uparrow}^{\dagger} \hat{c}_{(n+2)\downarrow}^{\dagger}
+\hat{c}_{(n+4)\uparrow}^{\dagger}\hat{c}_{(n+4)\downarrow}^{\dagger}
\hat{c}_{2\uparrow}^{\dagger} \hat{c}_{(n+3)\downarrow}^{\dagger}
+\hat{c}_{(n+5)\uparrow}^{\dagger}\hat{c}_{(n+5)\downarrow}^{\dagger}
\hat{c}_{3\uparrow}^{\dagger} \hat{c}_{(n+4)\downarrow}^{\dagger}
+\dots))
|0\rangle
\nonumber
\end{eqnarray}
Similar procedure applies for all basis wave vectors.
We mention that the so obtained basis wave functions are orthogonal.

Here we must note that because of the fixed conventions presented above, 
sometimes an additional negative sign arises in the process of writing the
mathematical expression corresponding to a basis wave vector component 
translated from the end to the beginning of the ladder in the presence of the 
periodic boundary conditions. 
For example, if we translate the vector 
$\hat c^{\dagger}_{1,\uparrow} \hat c^{\dagger}_{N/2,\uparrow}
\hat c^{\dagger}_{2,\downarrow} \hat c^{\dagger}_{3,\downarrow}|0\rangle$
by an elementary translation along the ladder, according to the fixed 
conventions one obtains 
$\hat c^{\dagger}_{2,\uparrow} \hat c^{\dagger}_{1,\uparrow}
\hat c^{\dagger}_{3,\downarrow} \hat c^{\dagger}_{4,\downarrow}|0\rangle=
- \hat c^{\dagger}_{1,\uparrow} \hat c^{\dagger}_{2,\uparrow}
\hat c^{\dagger}_{3,\downarrow} \hat c^{\dagger}_{4,\downarrow}|0\rangle$.

\subsection{The ground state wave function}

After the calculation presented above, we are in the possession of nine type of
orthogonal basis wave vectors $|A_i\rangle, |B_i\rangle, ..., 
|J_{i,j,k}\rangle$, enumerated together with their generating configuration
in Fig.2. Let as denote these basis wave vectors by $|O^{(m)}_{i,j,..}\rangle$,
$m=1,2,3, ..., 9$. Now one observes that by applying the Hamiltonian on a 
given $|O^{(m)}_{i,j,..}\rangle$ basis wave  vector with fixed $m$, we obtain
the result inside the $\{ |O^{(m)}_{i,j,..}\rangle \}$ set. Consequently,
nine explicitly given analytic linear equations form a closed system of 
equations, whose secular equation, by its minimum eigenvalue, contains the 
ground state at attractive $U$. The nine equations are exemplified in Appendix 
A and are available in their complete extent in Ref.[\cite{endre}].
The ground state nature of the minimum energy eigenstate has been 
tested by exact numerical diagonalizations taken on the full Hilbert space for
different $N$ values.

The fact that the analytic solution of the problem can be given in such a 
manner for arbitrary large ladder length is connected to the observation that 
with increasing $N$, the type of the particle configurations describing the 
system (see Fig.2), remains unchanged.
The deduction of the ground state itself from the system of equations
presented in Appendix A must be numerically given \cite{xx1}. Since the 
possible inter-particle distances (e.g. the possible values of the $i,j,..$ 
indices in $O^{(m)}_{i,j,..}$ at fixed $m$) depend on the $N$ value, the 
number of equations which must be numerically treated depends on $N$ in the 
frame of the same analytic expressions. For example, for the $m=1$ case we have
$1 < i \leq 1+N/4$, for the $m=2$ case we have $1 \leq i \leq 1+N/4$, etc.
The number of obtained equations $d_e$
is however significantly
 lower than $d_H$, the $c_{g}=d_H/d_e$ ratio being 
at least of order $10^2$ at intermediate $N\sim O(10)$ values. Increasing $N$, 
$c_{g}$ further increases.

\subsection{Application possibilities in other cases}

In fact, the deduced system of equations, based on symmetry 
properties, delimitates from the full Hilbert space a $d_e$ dimensional 
space region, inside of which the ground state is placed. The deduction of 
a such region is possible for other (non-disordered) models, and other 
particle numbers as well.
In order to do this, we mention that if the 
lattice sites are equivalent, the elementary translation of a 
particle configuration can be in principle given with a site independent 
multiplicative phase factor $exp(i \alpha_{trans})$. Furthermore, the rotation 
of a particle configuration along a symmetry axis can be given 
in principle with 
a multiplicative phase factor of the form $exp(i \alpha_{rot})$, both 
$\alpha_{trans}, \alpha_{rot}$ providing their contributions in the basis wave
vectors \cite{xx2}. In the described case, we have $\alpha_{trans} =
\alpha_{rot}=0$, but in other cases, the energy can be minimized in function of
these parameters.

In deducing the linear system of equations describing ${\cal{H}}_g$ in a new
case characterized by a new $\hat H$, one must start from a given basis wave 
vector (denoted by $|v_1\rangle$, for example). This is obtained
from a generating particle configuration, which is translated and 
rotated as specified above, all such obtained configurations being summed up.
From technical reasons, the first generating particle configuration must be 
such chosen to contain (for $1/2$ spin fermions) only double occupied sites
placed in nearest neighbor sites. Calculating now $\hat H |v_1\rangle$, 
the result becomes a linear combination containing new base vectors 
$|v_2\rangle, ..., |v_{n_1}\rangle$, holding the same symmetry properties, but
being related to new generating configurations. Continuing the procedure
by calculating $\hat H |v_2\rangle, |\hat H |v_3\rangle$, etc., since periodic
boundary conditions are used, the linear system of equations closes up. It is
even not important to know all distinct particle configuration possibilities,
since these are automatically generated by the $\hat H |v_i\rangle$ operation.

\section{Ground state properties}

By diagonalizing the system of equations presented in Appendix A and taking 
the minimum energy solution, one finds the ground state wave function
$|\Psi_g\rangle$. Using this, the complete quantum mechanical characterization
of the ground state can be given. In order to exemplify the results, we
present in (\ref{gs1},\ref{gs2}) explicit expressions containing the leading 
terms of the ground state wave function for two parameter values.
Even Appendix B shows that in the leading terms of 
the ground state wave function, the particles have the tendency to be placed 
in pairs, the pairs tending to occupy the highest possible distance between 
them. This is reflected as well in the density-density correlation function 
depicted in Fig.4c.

Ground state expectation values and correlation functions are exemplified
in Figs. 4-5. calculated for $N=28$, e.g. ladder containing 14 rungs described
by periodic boundary conditions taken along the ladder.
The correlation functions are defined as
follows. The density-density correlation function has the expression
\begin{eqnarray}
C_n(r)=\frac{1}{N}\sum_{i=1}^N (\langle 
\hat{n}_i \hat{n}_{i+r} \rangle - 
\langle \hat{n}_i \rangle \langle \hat{n}_{i+r}\rangle ) 
\label{cor1}
\end{eqnarray}
where $\hat{n}_i=\hat{n}_{i \uparrow}+\hat{n}_{i \downarrow}$, 
$\hat{n}_{i, \sigma}=\hat{c}_{i \sigma}^{\dagger}\hat{c}_{i \sigma}$.
The spin correlations are studied via
\begin{eqnarray}
C_{S^z}(r)=\frac{1}{N}\sum_{i=1}^N (\langle 
\hat{S}^z_i \hat{S}^z_{i+r} \rangle 
- \langle \hat{S}^z_i \rangle \langle \hat{S}^z_{i+r}\rangle ) ,
\label{corr2}
\end{eqnarray}
where $\hat{S}^z = (1/2) (\hat n_{i,\uparrow}-\hat n_{i,\downarrow})$.
The superconducting pairing s-wave \cite{swave} correlation function is 
\begin{eqnarray}
C_{sw}(r)=\frac{1}{N}\sum_{i=1}^N (\langle 
\hat{c}_{i \uparrow}^{\dagger}\hat{c}_{i \downarrow}^{\dagger}
\hat{c}_{(i+r) \downarrow}\hat{c}_{(i+r) \uparrow} \rangle - 
\langle \hat{c}_{i \uparrow}^{\dagger}\hat{c}_{(i+r) \uparrow} \rangle
\langle \hat{c}_{i \downarrow}^{\dagger}\hat{c}_{(i+r) \downarrow} \rangle), 
\label{cor3}
\end{eqnarray}
and the superconducting pairing d-wave \cite{dwave} 
correlations are studied via
\begin{eqnarray}
C_{dw}(r)= \frac{1}{N} \sum_{i=1}^N 
\langle \hat{\Delta}^{\dagger}(i+r) \hat{\Delta}(i) \rangle  
\label{cor4}
\end{eqnarray}
where $\hat{\Delta}(i)= (\hat{c}_{i_2 \downarrow}\hat{c}_{i_1 \uparrow}-
\hat{c}_{i_2 \uparrow}\hat{c}_{i_1 \downarrow})$. The $i$ in
$\hat \Delta (i)$ denotes a rung connecting the lattice sites $i_1,i_2$. 
The $r$ values inside the figures are given in lattice constant units.

Fig.4a presents the ground state energy and the potential energy in 
$t_{\parallel}$ units in function of $u=|U/t_{\parallel}|$ at $t_{\parallel}=
t_{\perp}$. Fig.4b shows that the spin-spin correlations are exponentially 
decreasing, the decrease rate in the $exp(-r/\xi)$ being of the form
$1/\xi= 0.34 + 0.78 \sqrt{|u|}$. 
The density-density correlations depicted in Fig.4c show that the 
particles tend to occupy opposite positions in the ladder closed by periodic
boundary conditions.

In Fig.5 the behavior of the superconducting correlation functions is 
presented. In these plots $u=U/t_{\parallel}$ holds. The correlations in
Fig.5 are decreasing with $r$, and for $s$-wave case slightly increase by 
increasing the attractive $U$ value. 
Fig.5c further shows that the decrease of the inter-leg 
hopping amplitude at fixed on-site interaction is detrimental to $d$-wave
pairing correlations. Similar behavior has been found also by others 
\cite{swave}.

\section{Summary and Conclusions}

We describe a procedure which allows the exact deduction of ground state
wave functions for few particles in lattice models. The main result of our 
paper is that indeed, a such type of analytic description can be done.
In the case of 
an arbitrary large two leg Hubbard ladder taken with periodic boundary 
conditions and containing four electrons, presented in details,
the method leads for the singlet 
state to nine analytic linear and coupled closed system of equations,
whose secular equation, through its minimum
eigenvalue solution, provides the ground state wave function and ground state
energy. The procedure is based on a ${\bf r}$-space representation of the wave
functions and properly constructed symmetry adapted orthogonal
basis wave vectors. These
are obtained from generating particle configurations translated and rotated
in the lattice and finally added. The linear system of equations is obtained 
by applying the Hamiltonian on the deduced basis wave vectors. The procedure 
can be applied for other systems as well.

The fact that the analytic structure of the ground state becomes visible by 
the use of the method underlines that the presented procedure contributes not
only to the understanding of theoretical aspects related to exact descriptions,
or development possibilities of new numerical approximation schemes, but has
implications on a broad spectrum of subfields related to technological
developments placed in between nano-devices and quantum computation, where
the exact knowledge of the behavior of a small number of quantum mechanical 
particles plays a main role.  

\acknowledgments

This work was supported by the Hungarian Scientific Research 
Fund through contract OTKA-T-037212. The numerical calculations have been 
done at the Supercomputing Lab. of the Faculty of Natural Sciences, 
Univ. of Debrecen, supported by OTKA-M-041537.

\appendix

\section{The linear system of equations containing the ground state.}
\def\theequation{{\thesection}\arabic{equation}}
This Appendix presents the nine analytic equations describing the action of 
the Hamiltonian on the basis wave vectors.

The first two equations are devoted to the $|A_i\rangle, |B_i\rangle$
species containing only (two) doubly occupied sites.  
\begin{eqnarray}
&&\hat{H}|A_i \rangle=2u|A_i \rangle - t_{\bot}|D_{i,i} \rangle - 
I_{i>2}|C_{i-1,i} \rangle- I_{i \le \frac{n}{2}}|C_{i,i+1} \rangle\:,
\nonumber\\
&&\hat{H}|B_i \rangle=2u|B_i \rangle - t_{\bot}I_{i>1}|D_{i,i} \rangle 
- I_{i>1}|E_{i-1,i} \rangle
-I_{i \le \frac{n}{2}}|E_{i,i+1} \rangle\:,
\nonumber
\end{eqnarray}
where $I_{K}=1$ if the statement $K$ is true, and $I_{K}=0$ otherwise. 

The following three equations describe the action of $\hat H$ on the basis
wave vectors containing only one doubly occupied site ($|C_{i,j}\rangle,
|D_{i,j}\rangle, |E_{i,j}\rangle$) as follows
\begin{eqnarray}
&&\hat{H}|C_{i,j} \rangle =
\nonumber\\
&&u|C_{i,j} \rangle -4\delta_{j,i+1}|A_i \rangle - 
4 \delta_{j,i+1}(1+\delta_{i,\frac{n}{2}})|A_{i+1} \rangle
-(1-\delta_{i,2})|C_{i-1,j} \rangle
\nonumber\\
&& - (1-\delta_{j,i+1}-\delta_{i,2}\delta_{j,\frac{n+i}{2}+1})|C_{i,j-1} 
\rangle
\nonumber\\
&&-(1-\delta_{i,2}\delta_{j,\frac{n+i}{2}}+\delta_{j,n-i+1}-
\delta_{j,n-i+2})|C_{i,j+1} \rangle
\nonumber\\
&&
-(1-\delta_{j,i+1})(1+\delta_{j,n-i+1}-\delta_{j,n-i+2})|C_{i+1,j} \rangle
\nonumber\\
&&+\delta_{i,2}(1+\delta_{j,3})(1-\delta_{j,\frac{n+i}{2}}-
\delta_{j,\frac{n+i}{2}+1})|C_{2,n-j+3} \rangle  
\nonumber\\
&&-t_{\bot}|D_{i,j} \rangle -t_{\bot}\cdot \left\{
\begin{array}{ll}
I_{j\le \frac{n}{2}+1} |D_{j,i} \rangle\\
I_{j> \frac{n}{2}+1}(1-\delta_{j,n-i+2}) |D_{n-j+2, n-i+2} \rangle
\end{array}\right \}
\nonumber\\
&&-(1-\delta_{i,2})(1+\delta_{j,i+1})|F_{i-1,i,n-j+i+1} \rangle
\nonumber\\
&&+\left\{\begin{array}{ll}
(1-\delta_{i,2})(1+\delta_{j,i+1})(1-\delta_{i,3}I_{j\ge \frac{n+i+1}{2}}) 
|F_{i,2,j} \rangle\\
-\delta_{i,3}I_{j> \frac{n+i+1}{2}} 
|F_{i,2,n-j+i+1} \rangle\\
\end{array}\right \}
\nonumber\\
&&-(1+\delta_{j,i+1})(1+\delta_{j,n-i+1}-\delta_{j,n-i+2})|F_{i,i+1,n-j+i+1} 
\rangle
\nonumber\\
&&+
(1+\delta_{j,i+1})(1+\delta_{j,n-i+1}-\delta_{j,n-i+2})
\times
\nonumber\\
&&\times
 \left\{
\begin{array}{ll}
(1-\delta_{i,2}I_{j\ge \frac{n+i}{2}}) |F_{i+1,2,j+1} \rangle\\
-\delta_{i,2}I_{j > \frac{n+i}{2}}|F_{i+1,2, n-j+i+1} \rangle
\end{array}\right \}
+t_{\bot}\cdot \left\{
\begin{array}{ll}
-I_{j\le \frac{n+i+1}{2}} |G_{i,j,1} \rangle\\
I_{j > \frac{n+i+1}{2}}|G_{i,n-j+i+1,i} \rangle
\end{array}\right \}
\nonumber\\
&&+t_{\bot}(1-\delta_{j,n-i+2})
\cdot \left\{
\begin{array}{ll}
I_{j\le \frac{n}{2}+1} 
\cdot \left\{
\begin{array}{ll}
I_{j < 2i-1} |G_{j,j-i+1,j} \rangle\\
-I_{j \ge 2i-1}|G_{j,i,1} \rangle
\end{array}\right \}
\\I_{j > \frac{n}{2}+1} 
\cdot \left\{
\begin{array}{ll}
-I_{j \le 2i-1}|G_{n-j+2,n-i+2,1} \rangle\\
I_{j > 2i-1} |G_{n-j+2,n-j+i+1,n-j+2} \rangle
\end{array}\right \}
\end{array}\right \} .
\nonumber
\end{eqnarray}
\begin{eqnarray}
&&\hat{H}|D_{i,j} \rangle =
\nonumber\\
&&u|D_{i,j} \rangle 
-4t_{\bot}\delta_{j,i}(|A_i \rangle+|B_i \rangle)
\nonumber\\
&&
-t_{\bot}(1-\delta_{j,1}-\delta_{j,i}+\delta_{j,n-i+2})\cdot
\left\{
\begin{array}{ll}
I_{j\le n-i+2}I_{j<i}(|C_{j,i} \rangle+ |E_{j,i} \rangle)   \\
 I_{n-i+2\ge j>i} (|C_{i,j} \rangle+|E_{i,j} \rangle)\\
I_{j>n-i+2}(|C_{n-j+2,n-i+2} \rangle+|E_{n-j+2,n-i+2} \rangle)
\end{array}\right \}
\nonumber\\
&&
+\left\{
\begin{array}{ll}
-(1-\delta_{i,2}) |D_{i-1,j} \rangle\\
\delta_{i,2}(1-\delta_{j,1}-\delta_{j,2}-\delta_{j,\frac{n+i}{2}}-
\delta_{j,\frac{n+i}{2}+1})|D_{i,n+i-j+1} \rangle
\end{array}\right \}
\nonumber\\
&&
-\left\{
\begin{array}{ll}
I_{j>1}[1-\delta_{i,2}(\delta_{j,2}+\delta_{j,\frac{n+i}{2}+1})+
\delta_{i,\frac{n}{2}+1}\delta_{j,2}] |D_{i,j-1} \rangle\\
\delta_{j,1}(1-\delta_{i,\frac{n}{2}+1})
|D_{i,n} \rangle
\end{array}\right \}
\nonumber\\
&&
-\left\{
\begin{array}{ll}
[1-\delta_{i,2}(\delta_{j,1}+\delta_{j,\frac{n+i}{2}})-\delta_{j,n}+
\delta_{i,\frac{n}{2}+1}
(\delta_{j,\frac{n}{2}}-\delta_{j,\frac{n}{2}+1})] |D_{i,j+1} \rangle\\
\delta_{j,n} |D_{i,1} \rangle
\end{array}\right \}
\nonumber\\
&&
-\left\{
\begin{array}{ll}
[I_{i<\frac{n}{2}}+\delta_{i,\frac{n}{2}}(I_{j<\frac{n}{2}+1}+\delta_{j,1}+
2\delta_{j,\frac{n}{2}+1})] |D_{i+1,j} \rangle \\
\big(\delta_{i,\frac{n}{2}}I_{j>\frac{n}{2}+1}+\delta_{i,\frac{n}{2}+1}
(1-\delta_{j,1}-\delta_{j,i})\big)
|D_{n-i+1,n-j+2} \rangle
\end{array}\right \}
\nonumber\\
&&
-(1-\delta_{i,2})\left\{
\begin{array}{ll}
I_{j \le i} |G_{i-1,i,i-j+1} \rangle \\
I_{j>i} |G_{i-1,i,n-j+i+1} \rangle
\end{array}\right \}
+\left\{
\begin{array}{ll}
[1-\delta_{i,2}-\delta_{i,3}(\delta_{j,2}+\delta_{j,3}+I_{j>\frac{n}{2}+1})] 
|G_{i,2,j} \rangle \\
-\delta_{i,3}\delta_{j,3}|G_{i,2,1} \rangle \\
-\delta_{i,3}I_{j>\frac{n}{2}+2} |G_{i,2,n-j+i+1} \rangle
\end{array}\right \}
\nonumber\\
&&
+\left\{
\begin{array}{ll}
(1-\delta_{i,\frac{n}{2}+1}) \left\{
\begin{array}{ll}
-I_{j\le i}|G_{i,i+1,i-j+1} \rangle \\
-I_{j> i}|G_{i,i+1,n+i-j+1} \rangle 
\end{array}\right \}
\\
\delta_{i,\frac{n}{2}+1}I_{1<j<\frac{n}{2}+1}(1-\delta_{n,4}\delta_{j,2}) 
|G_{i,2,n-j+2} \rangle
\end{array}\right \}
\nonumber\\
&&+
\left\{
\begin{array}{ll}
[1-\delta_{j,n}-\delta_{i,\frac{n}{2}+1}-\delta_{i,2}(\delta_{j,1}
+\delta_{j,2}+I_{j \ge \frac{n}{2}+1})] |G_{i+1,2,j+1} \rangle \\
\delta_{j,n} |G_{i+1,2,1} \rangle\\
-\delta_{i,\frac{n}{2}+1}I_{1<j<\frac{n}{2}+1}
|G_{\frac{n}{2},\frac{n}{2}+1,\frac{n}{2}+j} \rangle
\\
-\delta_{i,2}\left\{
\begin{array}{ll}
\delta_{j,2}|G_{i+1,2,1} \rangle \\
I_{\frac{n}{2}+1<j<n}|G_{i+1,2,n+i-j+1}  \rangle
\end{array}\right \}
\end{array}\right \}
\nonumber\\
&&-t_{\bot} \cdot
\left\{
\begin{array}{ll}
I_{1<j<i}|H_{j,j,n-i+j+1} \rangle \\
4\delta_{i,j}|H_{j,j,1} \rangle\\
(I_{i<j<n-i+2}+2\delta_{j,n-i+2})|H_{i,i,n-j+i+1} \rangle\\ 
I_{j>n-i+2}|H_{n-j+2,n-j+2,n+i-j+1} \rangle
\end{array}\right \}
+t_{\bot}\cdot \left\{
\begin{array}{ll}
(I_{1<j<i}+4\delta_{j,i})|J_{j,1,i} \rangle\\
(1-\delta_{j,i})(I_{i<j<n-i+2}+2\delta_{j,n-i+2})|J_{i,1,j} \rangle \\
I_{j>n-i+2}|J_{n-j+2,1,n-i+2} \rangle
\end{array}\right \} .
\nonumber
\end{eqnarray}
\begin{eqnarray}
&&\hat{H}|E_{i,j} \rangle = 
\nonumber\\
&&u|E_{i,j} \rangle 
- 4\delta_{j,i+1}[(1+\delta_{i,1})|B_i \rangle 
+(1+\delta_{i,\frac{n}{2}})|B_j \rangle] 
-t_{\bot} (1-\delta_{i,1}) \cdot \bigg[|D_{i,j} \rangle
\nonumber\\
&&+
 \left\{
\begin{array}{ll}
I_{j\le \frac{n}{2}+1} |D_{j,i} \rangle \\
I_{\frac{n}{2}+1 <j <n-i+2} |D_{n-j+2, n-i+2} \rangle
\end{array}\right \}
\bigg] - \left\{
\begin{array}{ll}
[1-\delta_{i,1}+\delta_{i,2}(\delta_{j,\frac{n}{2}+1}-I_{j> \frac{n}{2}+1}) ]
|E_{i-1,j} \rangle \\
\delta_{i,2}I_{j>\frac{n}{2}+1 } |E_{i-1, n-j+2} \rangle \\
\delta_{i,1}(1+\delta_{j,2}-\delta_{j,\frac{n}{2}+1}) |E_{2,n-j+2} \rangle
\end{array}\right \}
\nonumber\\
&&
-(1-\delta_{j,i+1})|E_{i,j-1} \rangle
-(1+\delta_{j,n-i+1}-\delta_{j,n-i+2}+\delta_{i,1}\delta_{j,\frac{n}{2}}-
\delta_{i,1}\delta_{j,\frac{n}{2}+1})|E_{i,j+1} \rangle
\nonumber\\
&&
-(1-\delta_{j,i+1})(1+\delta_{j,n-i+1}-\delta_{j,n-i+2})|E_{i+1,j} \rangle
+t_{\bot}(1-\delta_{i,1})\cdot \left\{
\begin{array}{ll}
-I_{j \le \frac{n+i+1}{2}} |G_{i,j,1} \rangle \\
I_{j>\frac{n+i+1}{2}} |G_{i,n-j+i+1,i} \rangle 
\end{array}\right \}
\nonumber\\
&&
+t_{\bot}(1-\delta_{i,1}-\delta_{j,n-i+2})\cdot \left\{
\begin{array}{ll}
I_{j \le \frac{n}{2}+1} 
\cdot \left\{
\begin{array}{ll}
I_{j < 2i-1} |G_{j,j-i+1,j} \rangle \\
-I_{j \ge 2i-1} |G_{j,i,1} \rangle 
\end{array}\right \}
\\
I_{j>\frac{n}{2}+1} 
 \cdot \left\{
\begin{array}{ll}
-I_{j \le 2i-1} |G_{n-j+2,n-i+2,1} \rangle \\
I_{j>2i-1} |G_{n-j+2,n-j+i+1,n-j+2} \rangle 
\end{array}\right \}
\end{array}\right \}
\nonumber\\
&&
+(1+\delta_{j,i+1})\cdot \left\{
\begin{array}{ll}
I_{i>2} |H_{i-1,n-j+i+1,i} \rangle \\
\delta_{i,2} |H_{1,2,n-j+i+1} \rangle \\
\delta_{i,1}\cdot \left\{
\begin{array}{ll}
2\delta_{j,2} |H_{2,2,1} \rangle \\
I_{\frac{n}{2}+1> j > 1} |H_{2,2,n-j+3} \rangle 
\end{array}\right \}
\end{array}\right \}
+[1+\delta_{j,i+1}(1+2\delta_{i,1})]|H_{i,2,j} \rangle
\nonumber\\
&&
+\left\{
\begin{array}{ll}
I_{i > 1}(1+\delta_{j,i+1})(1+\delta_{j,n-i+1}-\delta_{j,n-i+2}) 
|H_{i,n-j+i+1,i+1} \rangle \\
\delta_{i,1}(1+\delta_{j,i+1}-\delta_{j,\frac{n}{2}+1}) |H_{1,2,n-j+i+1} 
\rangle
\end{array}\right \}
\nonumber\\
&&
+(1+\delta_{j,i+1})(1+\delta_{j,n-i+1}-\delta_{j,n-i+2})|H_{i+1,2,j+1} 
\rangle .
\nonumber
\end{eqnarray}

The last four equations devoted to the base vectors
$|F_{i,j,k}\rangle, |G_{i,j,k}\rangle, |H_{i,j,k}\rangle, |J_{i,j,k}\rangle$
(not containing doubly occupied sites) can be find in Ref.[\cite{endre}].
  
\section{Exemplification for ground state wave functions}

We present below the leading terms of explicit ground state wave functions 
deduced for $N=28$, at $|U/t_{\parallel}|=3$. The ground state $|\Psi_g\rangle$
is normalized to unity, and contains ortho-normalized basis wave vectors.

For $t_{\bot}/t_{\parallel}=0.8$ one obtains for the ground state wave function
\begin{eqnarray}
&&|\Psi_g \rangle = 
\nonumber\\
&&0.181883|E_{7,8}\rangle +
0.181878|C_{7,8}\rangle +
0.175769|D_{7,7}\rangle +
0.169247|C_{6,7}\rangle \nonumber\\
&&+
0.169246|E_{6,7}\rangle +
0.157289|D_{6,6}\rangle +
0.145021|C_{5,6}\rangle +
0.145004|E_{5,6}\rangle \nonumber\\
&&+
0.138346|D_{8,7}\rangle +
0.138346|D_{7,8}\rangle +
0.128723|D_{6,7}\rangle +
0.128721|D_{7,6}\rangle \nonumber\\
&&+
0.12823|D_{5,5}\rangle +
0.111346|C_{4,5}\rangle +
0.111315|E_{4,5}\rangle +
0.110256|D_{5,6}\rangle \nonumber\\
&&+
0.110239|D_{6,5}\rangle +
0.10177|E_{6,8}\rangle +
0.101761|C_{6,8}\rangle +
0.097877|G_{7,8,1}\rangle \nonumber\\
&&
-0.0978768|G_{8,2,8}\rangle +
0.0931102|D_{7,9}\rangle +
0.0913856|D_{4,4}\rangle 
-0.0910721|G_{7,2,7}\rangle \nonumber\\
&&+
0.0910714|G_{6,7,1}\rangle +
0.0909671|C_{5,7}\rangle +
0.0909645|E_{5,7}\rangle +
0.0898481|D_{8,6}\rangle \nonumber\\
&&+
0.0898458|D_{6,8}\rangle +
0.0845566|D_{4,5}\rangle +
0.0844776|D_{5,4}\rangle +
0.0802931|D_{5,7}\rangle 
\nonumber\\
&&+ ...\,,
\label{gs1}
\end{eqnarray}
while for $t_{\bot}/t_{\parallel}=0.1$ one has
\begin{eqnarray}
&&|\Psi_g \rangle = 
\nonumber\\
&&
0.29866|E_{7,8}\rangle +
0.29365|E_{6,7}\rangle +
0.284039|E_{5,6}\rangle +
0.270726|E_{4,5}\rangle \nonumber\\
&&+
0.255311|E_{3,4}\rangle +
0.240496|E_{2,3}\rangle +
0.230556|E_{1,2}\rangle +
0.16658|C_{7,8}\rangle \nonumber\\
&&+
0.156952|C_{6,7}\rangle +
0.149053|E_{6,8}\rangle +
0.145317|E_{5,7}\rangle +
0.13942|E_{4,6}\rangle \nonumber\\
&&+
0.137923|C_{5,6}\rangle +
0.131927|E_{3,5}\rangle +
0.123835|E_{2,4}\rangle +
0.116908|E_{1,3}\rangle \nonumber\\
&&+
0.110056|C_{4,5}\rangle +
0.081715|C_{6,8}\rangle +
0.0755983|E_{6,9}\rangle +
0.0751106|H_{6,13,7}\rangle \nonumber\\
&&-
0.0750843|H_{7,2,9}\rangle +
0.0746358|C_{3,4}\rangle +
0.0744796|C_{5,7}\rangle +
0.0743108|E_{5,8}\rangle\nonumber\\
&& +
0.0740908|B_7\rangle -
0.073841|H_{5,13,6}\rangle -
0.0738214|H_{6,2,8}\rangle -
0.0722444|B_6\rangle \nonumber\\
&&+
0.0718117|E_{4,7}\rangle -
0.0713933|H_{5,2,7}\rangle -
0.0713921|H_{4,13,5}\rangle +
0.0693442|B_5\rangle 
\nonumber\\
&&+ ...
\label{gs2}
\end{eqnarray}



\newpage

\begin{figure}[h]
\centerline{\epsfbox{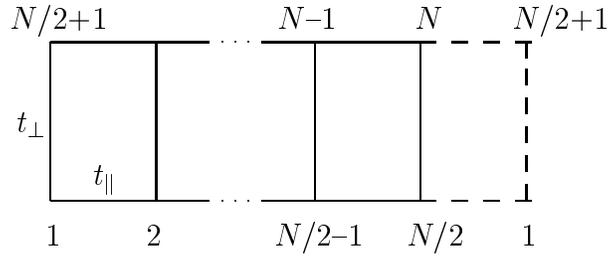}}
\caption{The numbering of the lattice sites for the two leg ladder
taken with periodic boundary conditions. $N$ denoting the number of lattice
sites is considered even. The $t_{\perp}$, $(t_{\parallel})$, denotes the
inter-leg, (intra-leg) hopping matrix element.}
\label{fig1}
\end{figure}

\newpage

\begin{figure}[h]
\centerline{\epsfbox{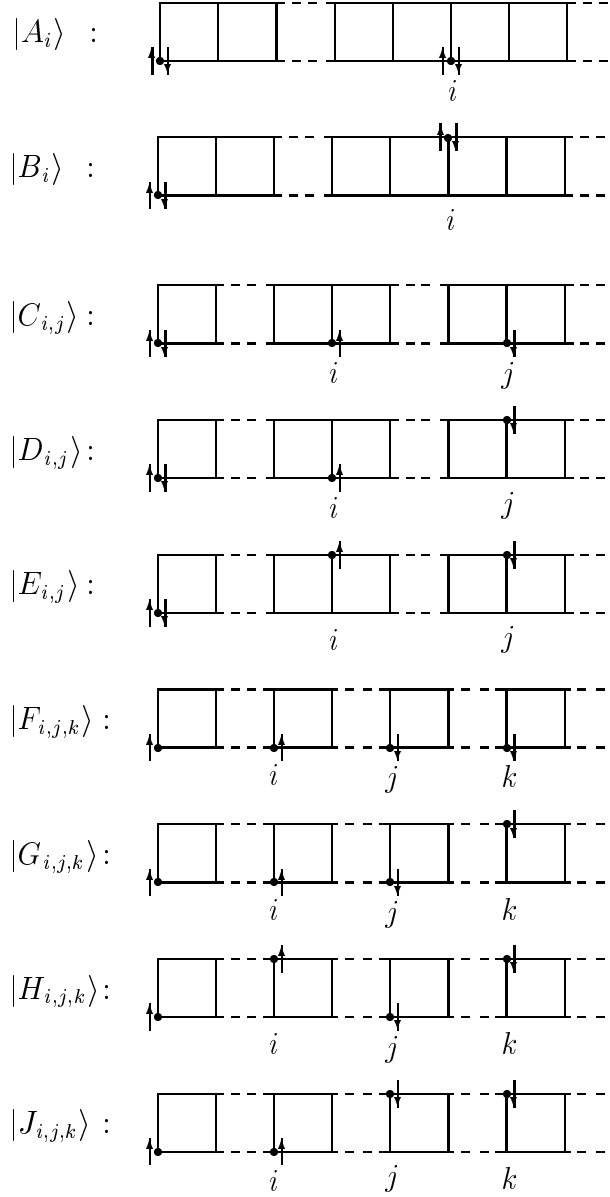}}
\caption{The different possible types of base vectors. We note that for the 
cases $C,E$ $i\ne j$, while for $F,J$ $j<k$ is considered, respectively. In 
the cases $F,G,H,J$, the double occupancy is forbidden.}
\label{fig2}
\end{figure}

\newpage

\begin{figure}[h]
\centerline{\epsfbox{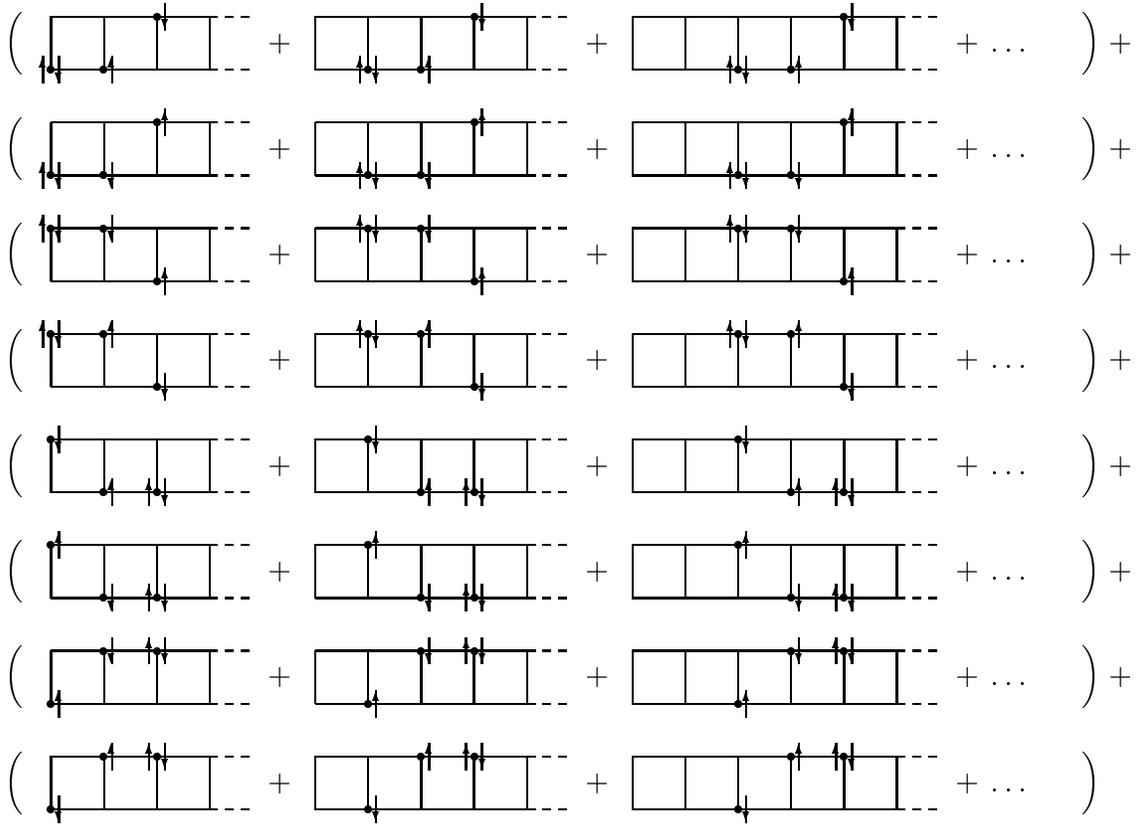}}
\caption{The structure of the $|D_{2,3}\rangle$ base vector.}
\label{fig3}
\end{figure}

\newpage

\begin{figure}[h]
\centerline{\epsfbox{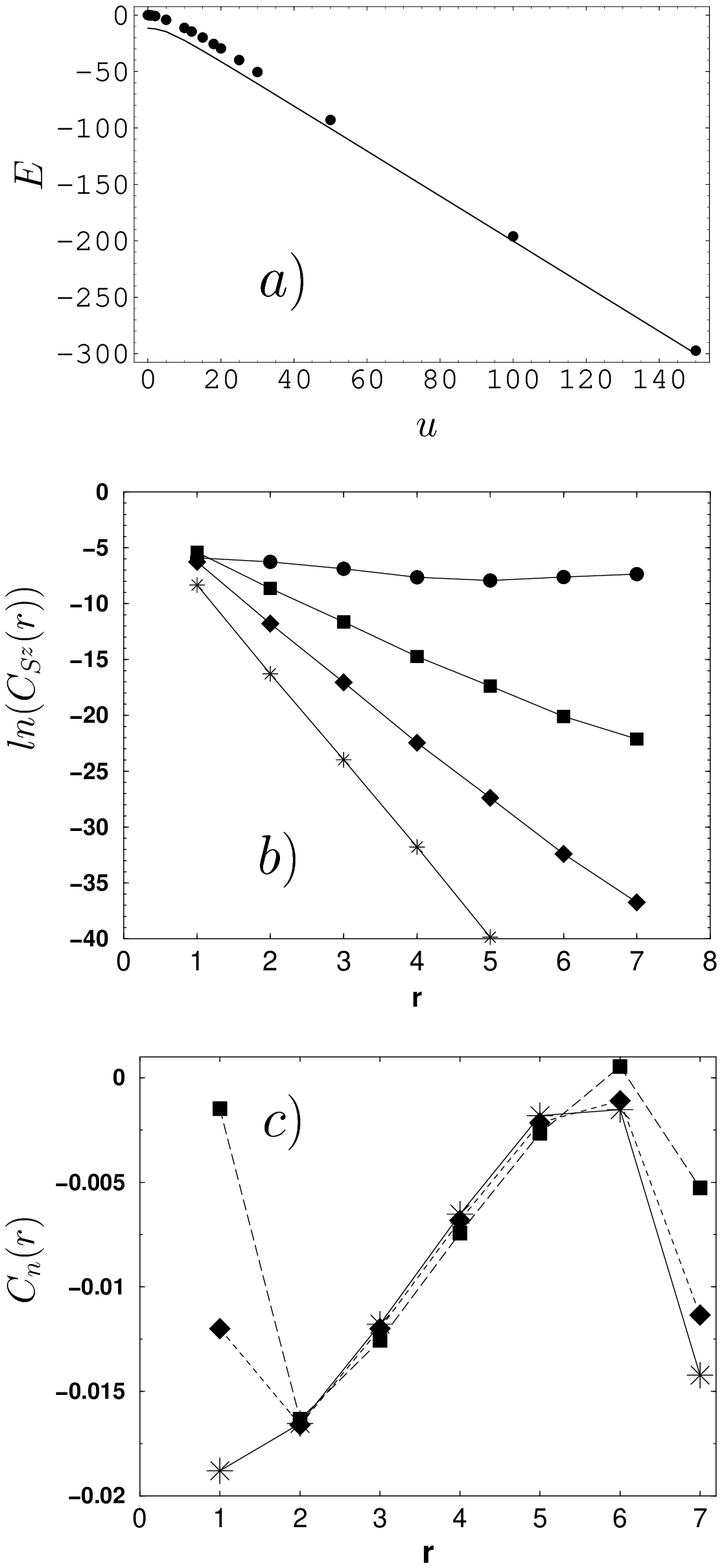}}
\caption{ }
\label{fig4}
\end{figure}

\newpage

Figure caption for Fig.4:

The properties of the ground state for $t_{\bot}=t_{\parallel}$.
(a) The dependence of the energy (in $t_{\parallel}$ units) on 
$u=U/t_{\parallel}$.
The continuous line is the total energy, while the dots indicate the 
potential energy.
(b) The logarithm of the same-leg $\hat{S}^z$-$\hat{S}^z$ 
correlation function for 
$u=0$ (dots, dot-dashed line),
$u=-10$ (squares, long dashed line),
$u=-30$ (diamonds, short dashed line),
$u=-100$ (stars, continuous line).
(c) The same-leg density-density correlation function for 
$u=0$ (dots, dot-dashed line),
$u=-10$ (squares, long dashed line),
$u=-30$ (diamonds, short dashed line),
$u=-100$ (stars, continuous line).

\newpage

\begin{figure}[h]
\centerline{\epsfbox{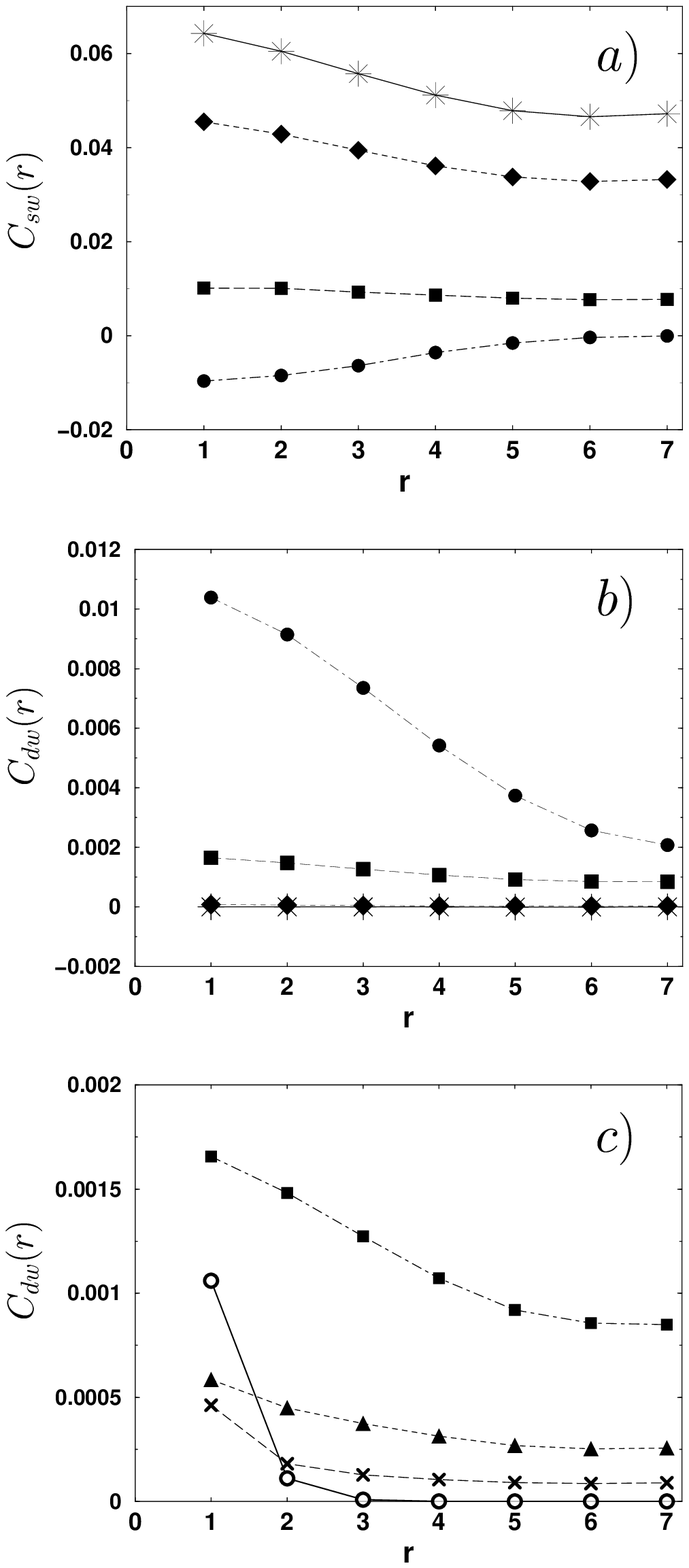}}
\caption{ }
\label{fig5}
\end{figure}

\newpage

Figure caption for Fig.5:

Superconducting ground state correlation functions. (a) 
The same-leg superconducting s-wave correlation function for 
$t_{\bot}=t_{\parallel}$ and
$u=0$ (dots, dot-dashed line),
$u=-10$ (squares, long dashed line),
$u=-30$ (diamonds, short dashed line),
$u=-100$ (stars, continuous line).
(b) The superconducting d-wave correlation function for 
$t_{\bot}=t_{\parallel}$ and
$u=0$ (dots, dot-dashed line),
$u=-10$ (squares, long dashed line),
$u=-30$ (diamonds, short dashed line),
$u=-100$ (stars, continuous line). We mention that the curves
corresponding to the last two $u$ values are almost superposed.
(c) Superconducting d-wave correlation function for $u=-10$ and 
$t=t_{\bot}/t_{\parallel}$ taken as 
$t=1$ (squares, dot-dashed line),
$t=0.5$ (triangles, short dashed line),
$t=0.3$ (X-s, long dashed line),
$t=0.01$ (circles, continuous line).


\begin{thebibliography}{dummy}
\bibitem{uu8}
We note that the here deduced ground states are entangled in the sense that 
cannot be factorized into a product of single-particle wave functions, although
the constituent particles are entirely distinct, see Ref.[\cite{uu6}].
\bibitem{xx1}
Such property is present also in the case of the Bethe Ansatz as well.
\bibitem{xx2}
The $\alpha_{trans}$ and $\alpha_{rot}$ angles can be connected to the
momentum and angular momentum values in the ground state.
\end{thebibliography}
\end{document}